\documentclass[a4paper,pdflatex,fleqn]{jpconf}
\usepackage{graphicx}
\pdfoutput=1
\begin{document}

\title{Experimental Status of Neutrino Physics}

\author{Fanny Dufour} 
\address{Section de Physique, Universit\'{e} de Gen\`{e}ve, 1205 Gen\`{e}ve, Switzerland}
\ead{fanny.dufour@unige.ch}

\author{David Wark}
\address{Imperial College London, Department of Physics, London, United Kingdom}
\ead{david.wark@stfc.ac.uk}

\begin{abstract}
After a fascinating phase of discoveries, neutrino physics still has a
few mysteries such as the absolute mass scale, the mass hierarchy, the
existence of CP violation in the lepton sector and the existence of
right-handed neutrinos. It is also entering a phase of precision
measurements. This is what motivates the NUFACT~11 conference which
prepares the future of long baseline neutrino experiments. In this
paper, we report the status of experimental neutrino physics. We focus
mainly on absolute mass measurements, oscillation parameters and
future plans for oscillation experiments~\cite{wark_nufact11}.
\end{abstract}

\section{Introduction}

Over the last 15 years experiments have demonstrated that neutrinos
oscillate and therefore mix and have masses, however few oscillation
parameters are well measured. In addition, their absolute mass
scale is still only described by upper limits. Finally, we do not
know the nature of neutrino masses. Therefore despite the tremendous
results of the last decade, there is still a vast open field with
many discoveries in the making. In this paper we present the status of
the absolute mass scale measurements and oscillation parameters
measurements, and describe several possible new experiments for
improving these measurements. This paper has been updated to take into
account the results presented at the Neutrino 2012 conference in June
2012 in Kyoto~\cite{nu12}.

\section{Measurements of the absolute neutrino mass}
Absolute neutrino masses can be measured in several
ways. Astrophysical neutrinos, kinematic limits and neutrinoless
double beta decay are all used to measure the absolute neutrino mass
scale or set a limit on it.

\subsection{Supernovae and cosmological constraints}
Supernovae are copious sources of neutrinos: by measuring time shifts,
it is in principle possible to measure neutrino masses down to 30~eV
as in the case of SN1987a~\cite{Arnett:1987iz}. Since neutrinos are so
numerous in the universe, even a tiny neutrino mass can have
cosmological implications. Current cosmological bounds are down around
$M_{\nu} < 0.17-0.33$~eV~\cite{bookneutrino,Spergel:2003cb} and new
data from the Planck observatory should be available
soon~\cite{Planck}. These results are quite model dependent and mostly
illustrate the sensitivity of structure formation to neutrinos
masses. They cannot replace direct laboratory experiments.

\subsection{Tritium beta decay and KATRIN}
Double beta decay experiments study the end of the beta decay spectrum
in order to measure $m^2_{\nu_e} = \sum_{i=1}^3 |U_{ei}|^2 m^2_i$. The
current upper electron-neutrino mass limit from beta decay experiments
is $m_{\nu_e}<2.3$~eV~\cite{Kraus:2004zw}. The next generation
experiment is KATRIN at Karlsruhe. It is sensitive down to
0.2~eV~\cite{katrin_talk,katrin_blois12}.

\begin{figure}[h]
\begin{center}
{\hbox{\hspace{0.0in}
    \includegraphics[width=18pc]{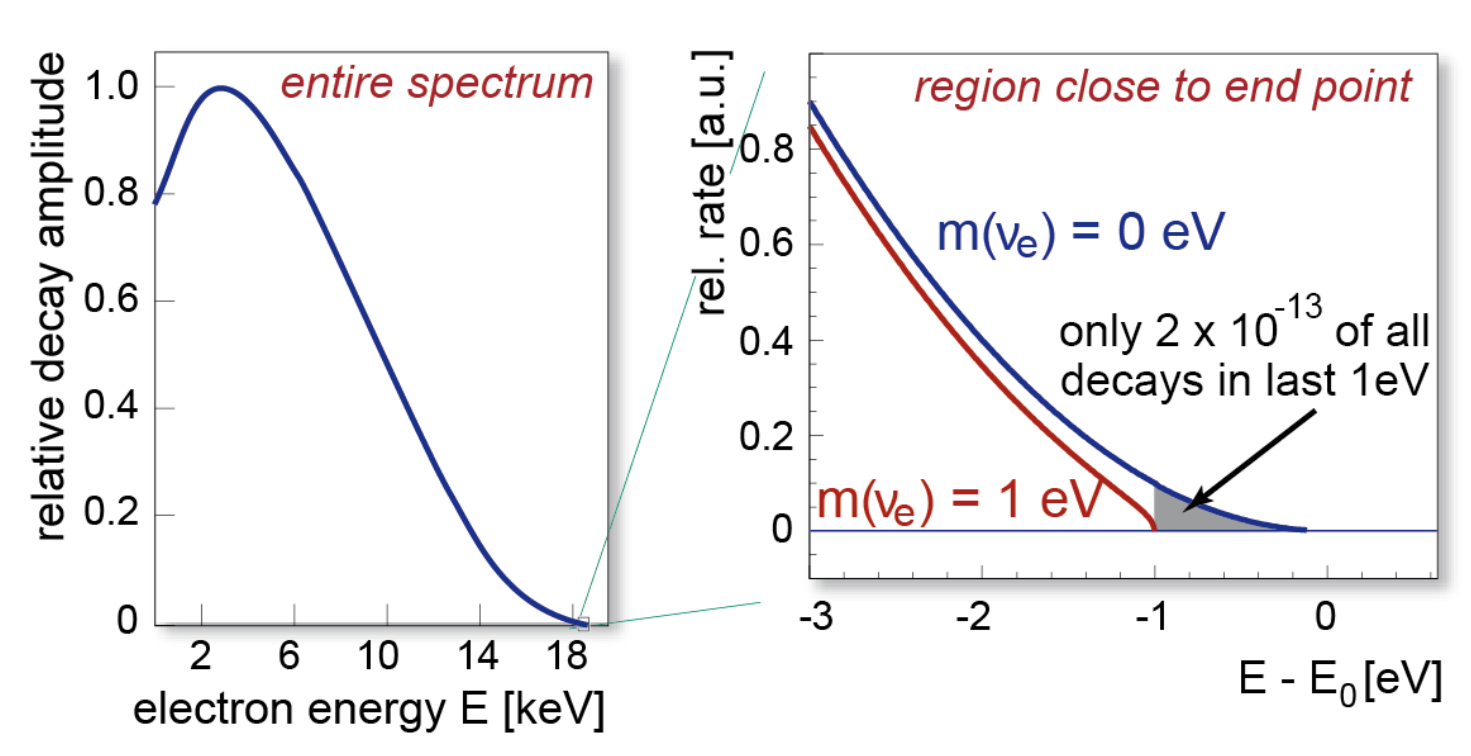}
\hspace{0.0in}
    \includegraphics[width=18pc]{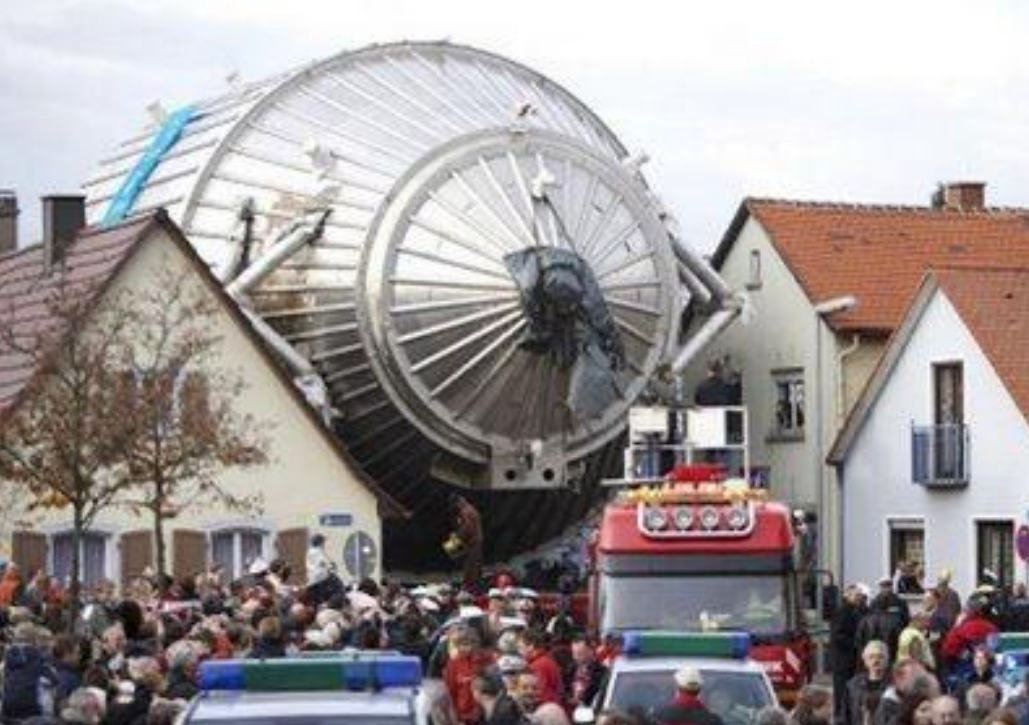}
}}
\end{center}
\vspace{-0.5cm}
\caption{\label{fig:tritium}{\bf Left}: Beta decay spectrum. {\bf
    Right}: KATRIN detector. \cite{katrin_talk}}
\end{figure}

\subsection{Neutrinoless double beta decay}
\begin{figure}[h]
\begin{center}
{\hbox{\hspace{0.0in}
    \includegraphics[height=12pc]{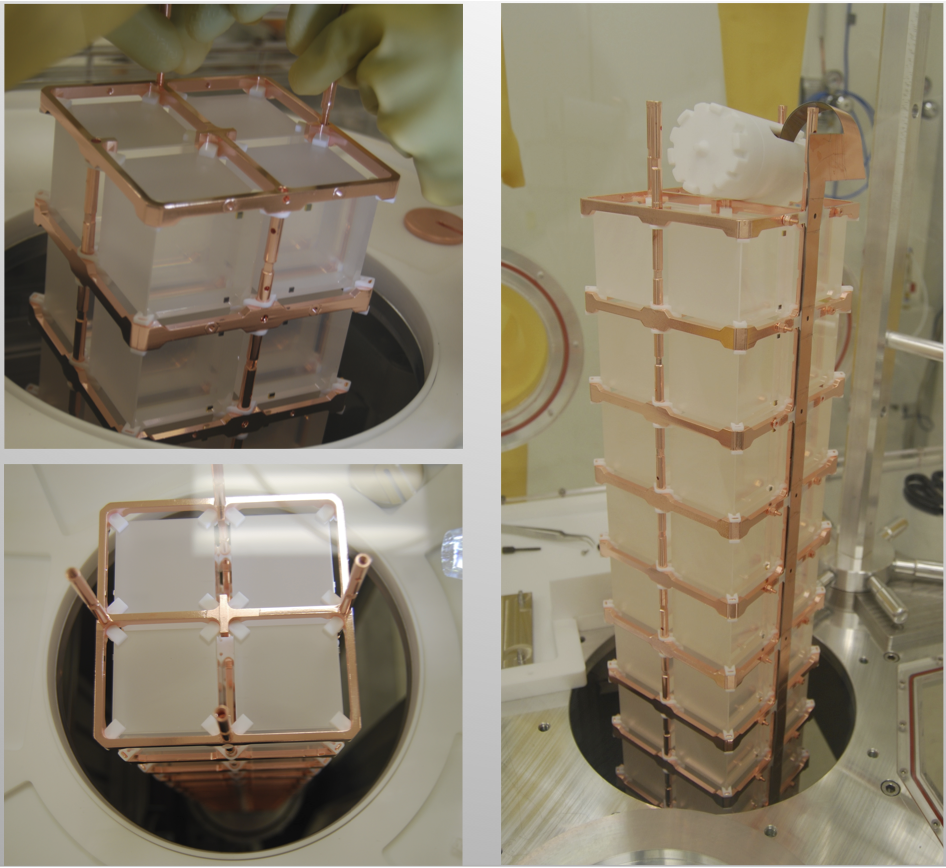}
\hspace{0.0in}
    \includegraphics[height=12pc]{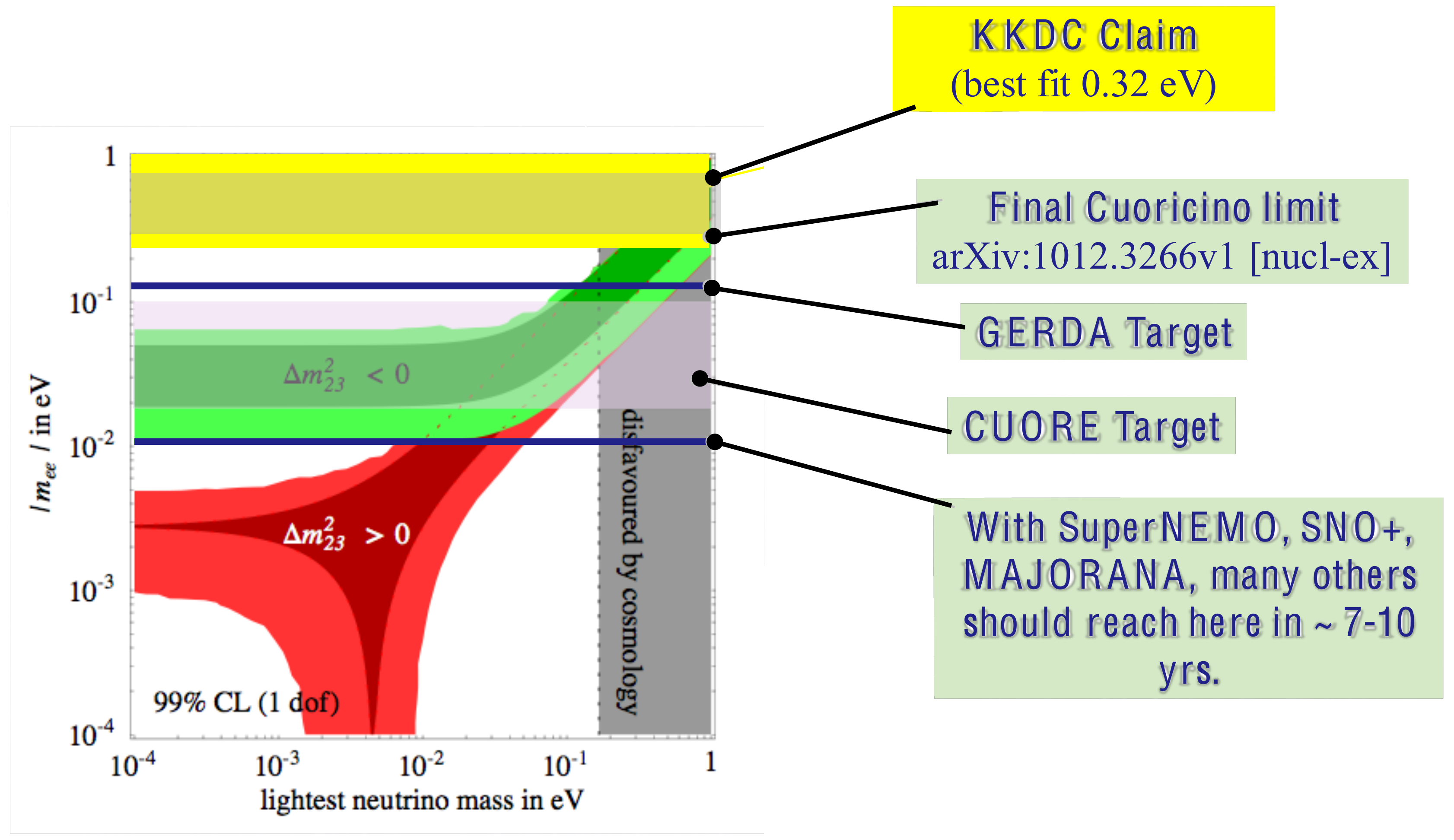}
}}
\end{center}
\vspace{-0.5cm}
\caption{\label{fig:onbb}{\bf Left}: CUORE-0 detector~\cite{cuore_nu12}. {\bf Right}:
  Neutrinoless double beta decay. Target sensitivity of several
  experiments~\cite{wark_nufact11}.}
\end{figure}

Neutrinoless double beta decay ($0\nu\beta\beta$) experiments are the
main method to investigate whether neutrinos have a Majorana mass term
which couples left-handed neutrinos to right-handed anti-neutrinos. It
is also an excellent tool to probe the absolute mass scale and, in
combination with oscillation experiments, gives a hint on the mass
hierarchy (Fig.~\ref{fig:onbb}).

A whole set of $0\nu\beta\beta$-decay experiments are currently
running or are about to start taking data: GERDA, CUORE-0, EXO and
KamLAND-Zen began in 2011. SNO+, Majorana and NEXT should start in
2013, and SuperNEMO, Lucifer, EXO-gas, XMASS and CUORE are scheduled
to start later. While we cannot report every result here, we report
that KamLAND-Zen found that $\langle m_{\beta\beta}\rangle<
0.26-0.54$~eV at 90\% C.L.~\cite{kamlandzen_nu12} and that EXO-200
found $ \langle m_{\beta\beta}\rangle< 0.14-0.38$~eV at the same
significance~\cite{Auger:2012ar}.

\section{Oscillation parameters measurements}

\begin{figure}[h]
\begin{center}
{\hbox{\hspace{0.0in}
    \includegraphics[height=11pc]{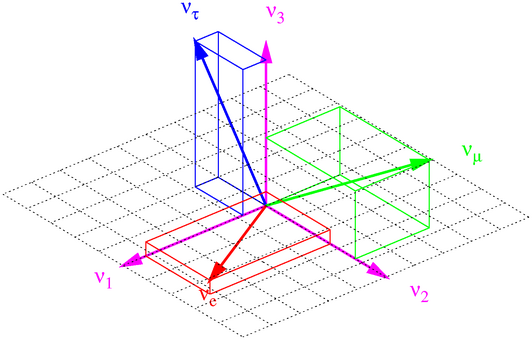}
\hspace{0.0in}
    \includegraphics[height=11pc]{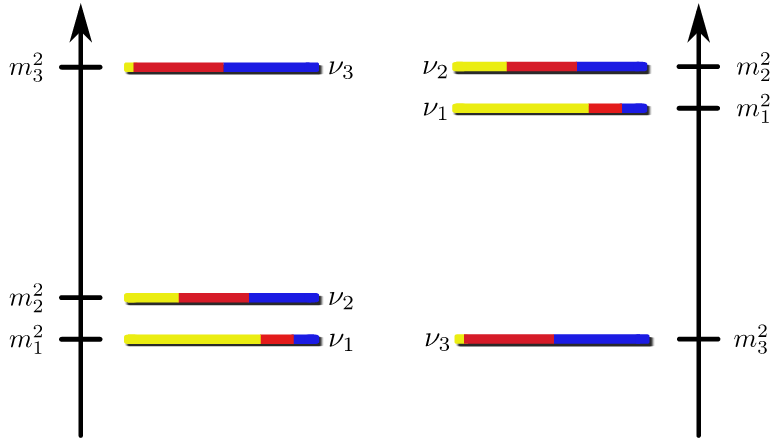}
}}
\end{center}
\vspace{-0.5cm}
\caption{\label{fig:nt}{\bf Left}: Neutrino mixing angles. {\bf Right}:
  Neutrino mass hierarchy (normal, inverted).}
\end{figure}

Since solar neutrino experiments revealed the famous solar neutrino
problem, we have known that neutrinos behave atypically. Since then
the fact neutrinos transmute and thus have mass has been established
by the Super-Kamiokande measurement of atmospheric
neutrinos~\cite{Ashie:2005ik} and by solar data from
SNO~\cite{Ahmad:2002jz}. Neutrino oscillations is governed by a $3
\times 3$ unitary mixing matrix and by mass differences. The mixing
angles (Fig.~~\ref{fig:nt}, left) have all been measured but even
though the two mass splittings have also been measured
(Fig.~\ref{fig:nt}, right), the mass hierarchy and the CP violating
phase are still unknown. In this section, we will present the current
status of the measurement of each oscillation parameter.
\vspace{-0.25cm}

\subsection{Measurement of $\theta_{13}$}

\begin{figure}[h]
\begin{center}
{\hbox{\hspace{0.0in}
    \includegraphics[width=18pc]{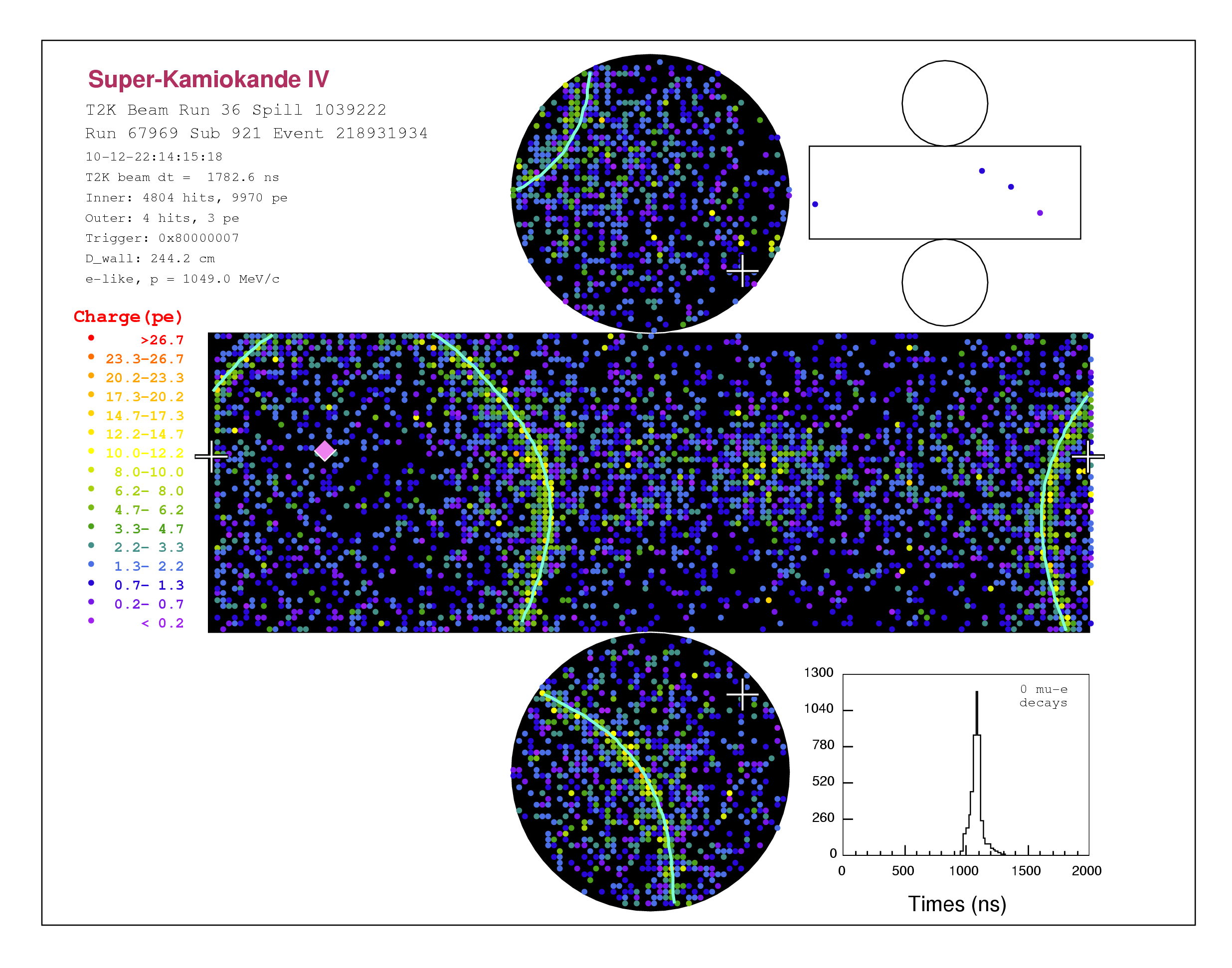}
\hspace{0.0in}
    \includegraphics[width=14.5pc]{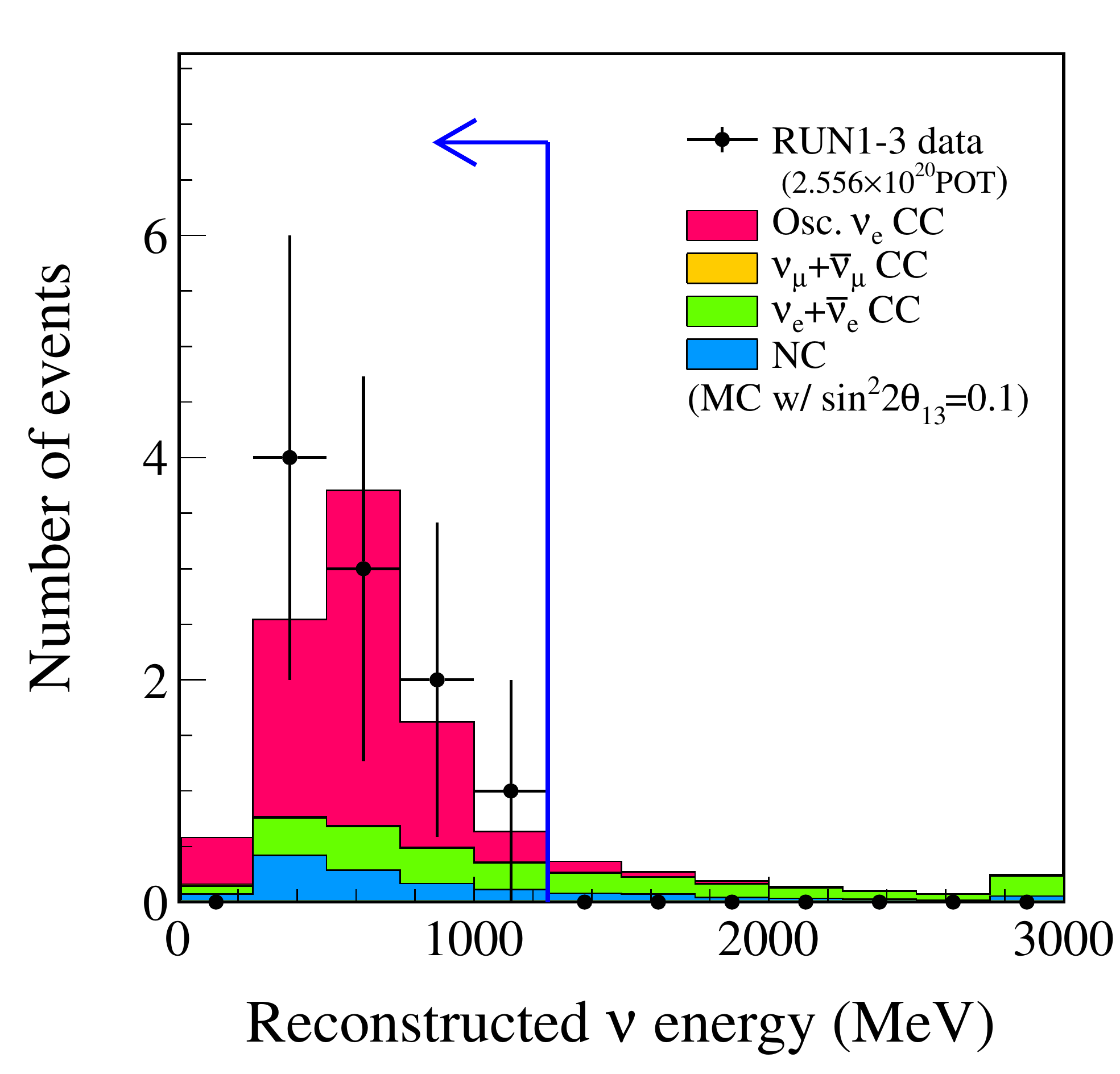}
}}
\end{center}
\vspace{-0.5cm}
\caption{\label{fig:t2k_nue}{\bf Left}: An electron neutrino event in
  SK. {\bf Right}: Reconstructed neutrino energy distribution for
  fully-contained, fiducial volume, single-ring e-like events with
  $\mathrm{Evis}>100~\mathrm{MeV}$, no decay-e and POLfit mass less
  than $105~\mathrm{MeV/c^2}$ in RUN1+2+3~\cite{T2K_nu12}.}
\end{figure}

Until last year, $\theta_{13}$ was the sole remaining unknown mixing
angle, the only knowledge of this parameter came from the Chooz
experiment and was an upper limit of 0.15 at
90\%~C.L.~\cite{Apollonio:2002gd}. A hint that $\theta_{13}$ was
non-zero did however come from global fit of available oscillation
data~\cite{fogli_nu12}. In June of 2011, T2K presented their first
electron neutrino appearance measurement, six events on a predicted
background of $1.5 \pm 0.3(syst.)$ indicating a non-zero value of
$\theta_{13}$ at 2.5$\sigma$~\cite{Abe:2011sj}. The analysis was not
strictly speaking blind but all the cuts for the electron neutrino
event selection in the Super-Kamiokande detector were decided well
before T2K even started taking data. At Neutrino 2012, all the data up
to May 15th 2012 were analyzed and presented~\cite{T2K_nu12}. Ten
events are observed in the far detector and $\theta_{13}$ is excluded
at $3.2\sigma$. At 90\%~C.L., the T2K results is $\sin^2 2\theta_{13}
=0.104 ^{+0.60}_{-0.45}$ for $\delta_{CP}=0$ and normal hierarchy. The
T2K selected $\nu_e$ candidates are shown in
Fig.~\ref{fig:t2k_nue}. Two weeks after T2K released their results,
the MINOS collaboration presented results which disfavour
$\theta_{13}=0$ at 89\% C.L. with 62 events on a background of $49.6
\pm 7.0(stat) \pm 2.7(syst.)$~\cite{Adamson:2011qu}. At Neutrino 2012,
MINOS disfavors $\theta_{13}=0$ at 96\% C.L. ~\cite{minos_nu12}.

\begin{figure}[h]
\begin{center}
{\hbox{\hspace{0.0in}
    \includegraphics[width=17pc]{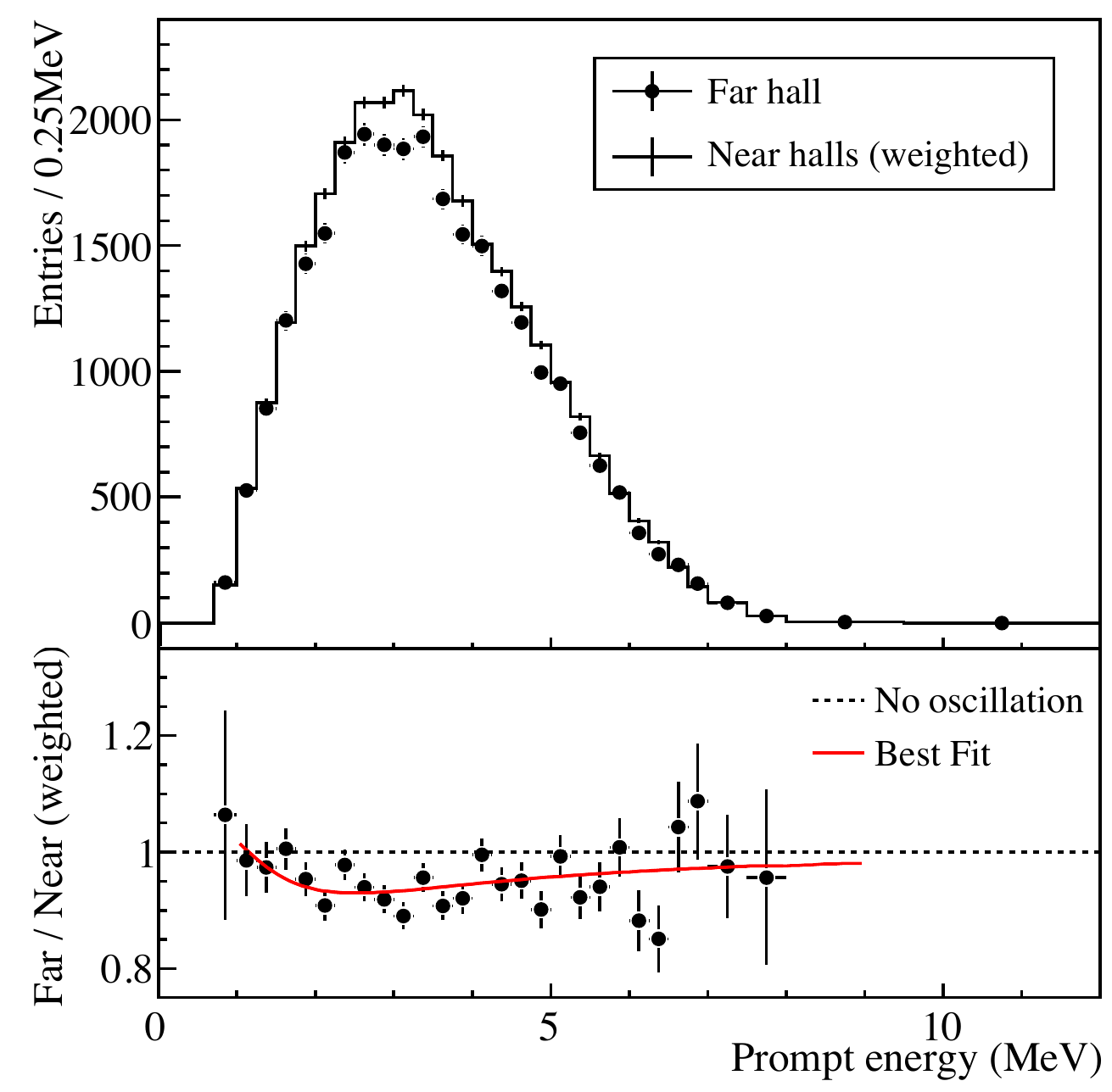}
\hspace{0.0in}
    \includegraphics[width=17pc]{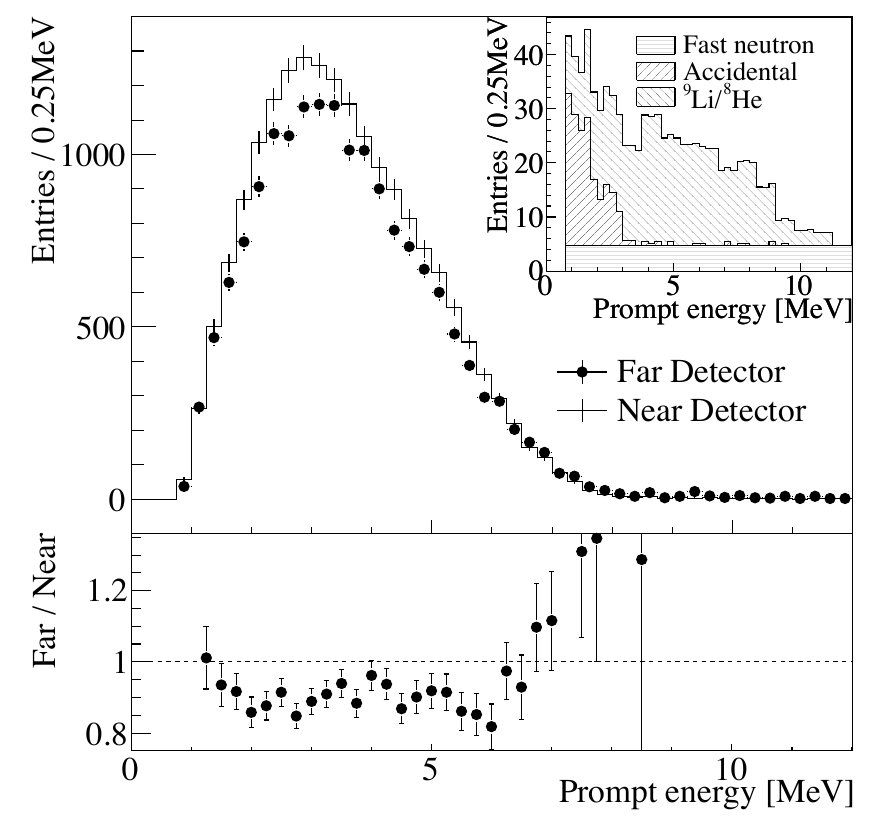}    
}}
\end{center}
\vspace{-0.5cm}
\caption{\label{fig:reactor_data} {\bf Left, Daya Bay data:} Top:
  Measured prompt energy spectrum of the far hall (sum of three ADs)
  compared with the no-oscillation prediction from the measurements of
  the two near halls. Spectra were background
  subtracted. Uncertainties are statistical only. Bottom: The ratio of
  measured and predicted no-oscillation spectra. The red curve is the
  best- fit solution with $\sin^2 2\theta_{13}= 0.089$ obtained from
  the rate-only analysis. The dashed line is the no-oscillation
  prediction~\cite{An:2012eh,dayabay_nu12}. {\bf Right, RENO data:}
  Observed spectrum of the prompt signals in the far detector compared
  with the non-oscillation predictions from the measurements in the
  near detector.  The backgrounds shown in the inset are subtracted
  for the far spectrum. The background fraction is $5.5\%$ ($2.7\%$)
  for far (near) detector. Errors are statistical uncertainties only.
  Bottom: The ratio of the measured spectrum of far detector to the
  non-oscillation prediction~\cite{Ahn:2012nd}.}
\end{figure}

Finally in March 2012, outstanding results came from the reactor
experiments Daya Bay~\cite{An:2012eh,dayabay_nu12} and
Reno~\cite{Ahn:2012nd}. The Daya Bay collaboration excluded
$\theta_{13} = 0 $ at $5.2 \sigma$~\cite{An:2012eh,dayabay_nu12},
while RENO excluded it at $4.9 \sigma$~\cite{Ahn:2012nd}. Their best
fit values are respectively $\sin^2 2\theta_{13} = 0.089 \pm
0.010(stat.)\pm 0.005(syst.)$ and $\sin^2 2\theta_{13} = 0.113 \pm
0.013(stat.)\pm 0.019(syst.)$ and their data is also shown in
Fig.~\ref{fig:reactor_data}. Double Chooz excluded $\theta_{13} = 0 $
at $3.1 \sigma$ and their best fit is $\sin^2 2\theta_{13} = 0.109 \pm
0.030(stat.)\pm 0.025(syst.)$~\cite{dchooz_nu12}.  All experiments are
still statistically dominated and the clearly demonstrate that
$\theta_{13}$ is greater than zero.
\vspace{-0.5cm}
\begin{figure}[h]
\begin{center}
{\hbox{\hspace{0.0in}
    \includegraphics[height=14pc]{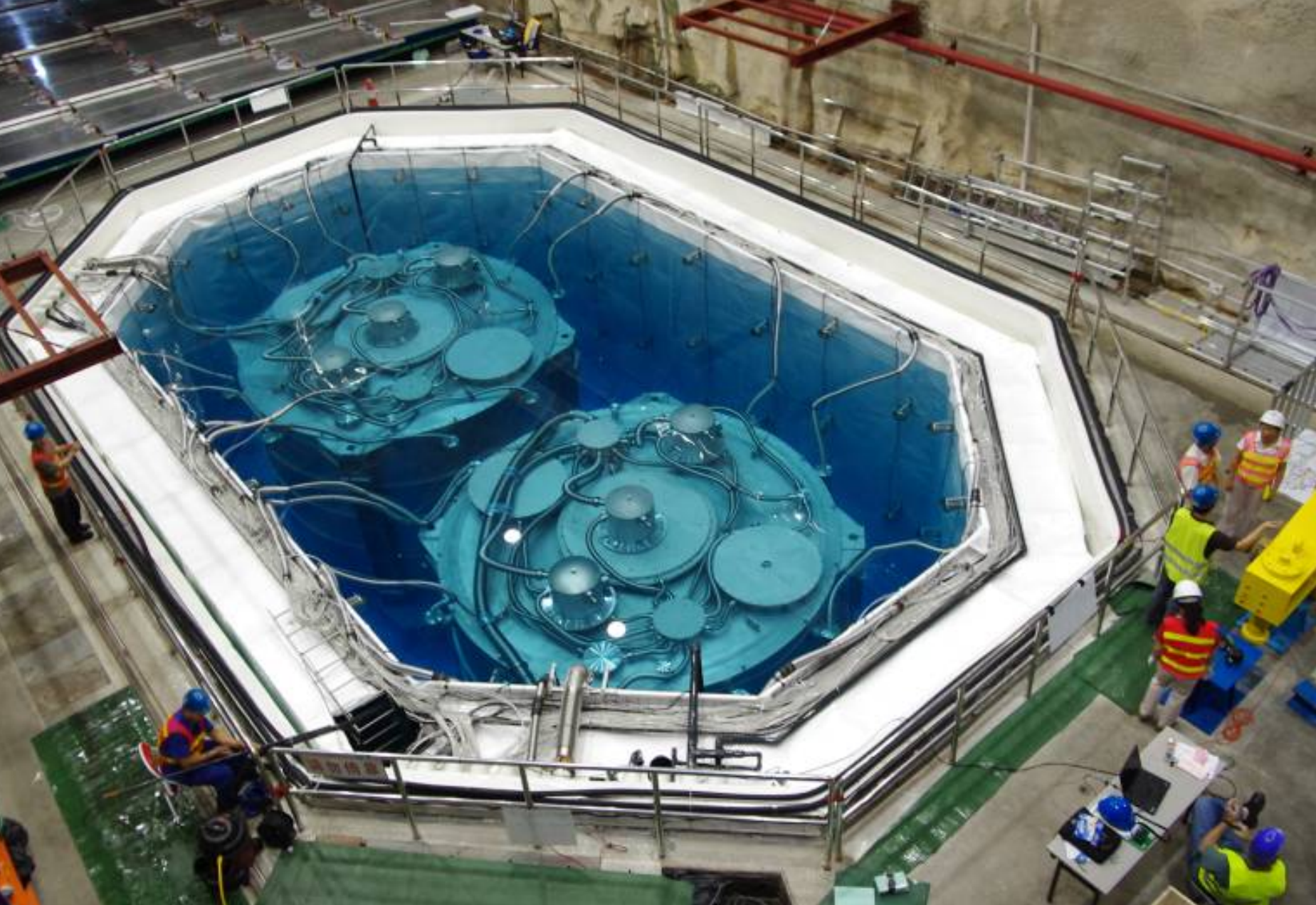}
\hspace{0.0in}
    \includegraphics[height=15pc]{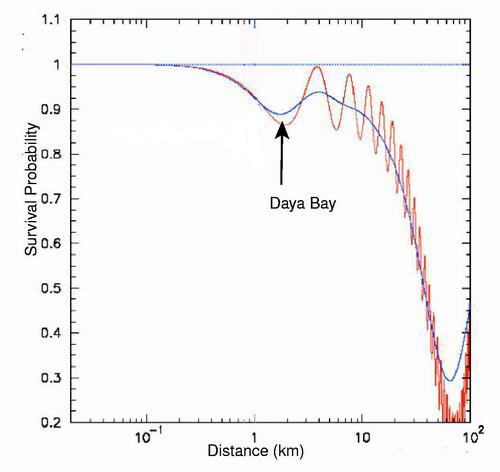}
}}
\end{center}
\vspace{-0.5cm}
\caption{\label{fig:t13_reactor}{\bf Left}: Picture of Daya Bay
  detectors~\cite{dayabay_nu12}. {\bf Right}: Survival probability of
  electron neutrino as a function of the distance.}
\end{figure}

Reactor experiments measure the disappearance of electron
anti-neutrinos to measure $\theta_{13}$ as shown in
Fig.~\ref{fig:t13_reactor} and Eq.~\ref{eq:disap}.  Current
accelerator experiments measure the appearance of electron neutrinos in
a muon-neutrino beam (Eq.~\ref{eq:app}). In addition the accelerator
measurement depends on the value of the CP phase $\delta$ while the
reactor measurement does not. Because these experiments are
fundamentally different it is very interesting to continue both in
parallel.

\begin {equation}
P_{\bar{\nu}_e\rightarrow \bar{\nu}_e} \approx 1 - \sin^2 2\theta_{13} \sin^2(1.267 \Delta m^2_{13} L/E)
\label{eq:disap}
\end{equation}
\newpage

\begin{eqnarray}
P[\nu_{\mu} &\rightarrow& \nu_e] \nonumber\\
&=& sin^2 2 \theta_{13} s^2_{23} \sin^2 \left( \frac{\Delta m^2_{31} L }{4E} \right) - \frac{1}{2} s^2_{12} \sin^2 2\theta_{13} 
s^2_{23} \left(\frac{\Delta m^2_{21}L}{2E}\right)\sin \left(\frac{\Delta m^2_{31}L}{2E}\right) \nonumber\\
&+& 2J_r \cos \delta \left(\frac{\Delta m^2_{21}L}{2E}\right) \sin \left(\frac{\Delta m^2_{31}L}{2E}\right) - 4J_r \sin \delta 
 \left(\frac{\Delta m^2_{21}L}{2E}\right) \sin^2 \left(\frac{\Delta m^2_{31}L}{4E}\right)  \nonumber\\
&+& \cos 2 \theta_{13} sin^2 2 \theta_{13} s^2_{23} \left( \frac {4Ea(x)}{\Delta m^2_{31}} \right)
\sin ^2 \left( \frac{\Delta m^2_{31} L }{4E}\right) \nonumber\\
&-& \frac{a(x) L }{2} \sin^2 2 \theta_{13} \cos 2 \theta_{13} s^2_{23} \sin \left( \frac{\Delta m^2_{31} L }{2E} \right)
+ c^2_{23} \sin^2 2 \theta_{12} \left (\frac{\Delta m^2_{21} L } {4E}\right )^2, \nonumber \\
\label{eq:app}
\end{eqnarray}

where $a(x) = \sqrt{2} G_F N_e (x)$, $G_F$ is the Fermi constant,
$N_e (x)$ is the electron number density at $x$ in the earth, $J_r =
c_{12} s_{12} c^2_{12} s_{13} c_{23} s_{23}$ is the Jarlslog determinant.

\subsection{Measurement of the solar parameters: $\Delta m^2_{21}$ and  $\theta_{12}$}

The solar parameters are now well constrained by both solar and
reactor data, and the newly measured value of $\theta_{13}$ further
reduces the uncertainty of this measurement. A very good summary of
current data is presented by T. Schwetz~\cite{Schwetz:2011qt} and the
current values of these parameters are given below.

$$ \Delta m^2_{21} = 7.59^{+0.20}_{-0.18} \times 10^{-5}
\mathrm{~eV^2} \mathrm{~~ and ~~} \sin^2 \theta_{12} =
0.312^{+0.017}_{-0.015} $$

The values of $\Delta m^2_{21}$ and $\sin^2 \theta_{12}$ have been
obtained using the KamLAND neutrino data and the Super-Kamiokande and
SNO solar data.

%
%
%
%
%
\subsection{Measurement of atmospheric parameters: $|\Delta m^2_{31}|$ and $\theta_{23}$}

The atmospheric parameters have also been well measured by the
Super-Kamiokande atmospheric data and by the MINOS experiment. Their
results have also been summarized by T.Schwetz~\cite{Schwetz:2011qt}
and are given below.

$$
 |\Delta m^2_{31}| = \left\{ 
 \begin {array}{ll}
   2.45 \pm 0.09 &\times 10^{-3} \mathrm{~eV^2~~(NH)} \\
   2.34 ^{+0.10}_{-0.09} & \times 10^{-3} \mathrm{~eV^2~~(IH)}\\
\end{array}
\right.  \mathrm{~~ and ~~ } \sin^2 \theta_{23} = 0.51\pm 0.06 $$

\vspace{-0.4cm}
\begin{figure}[h]
\begin{center}
\includegraphics[width=23pc]{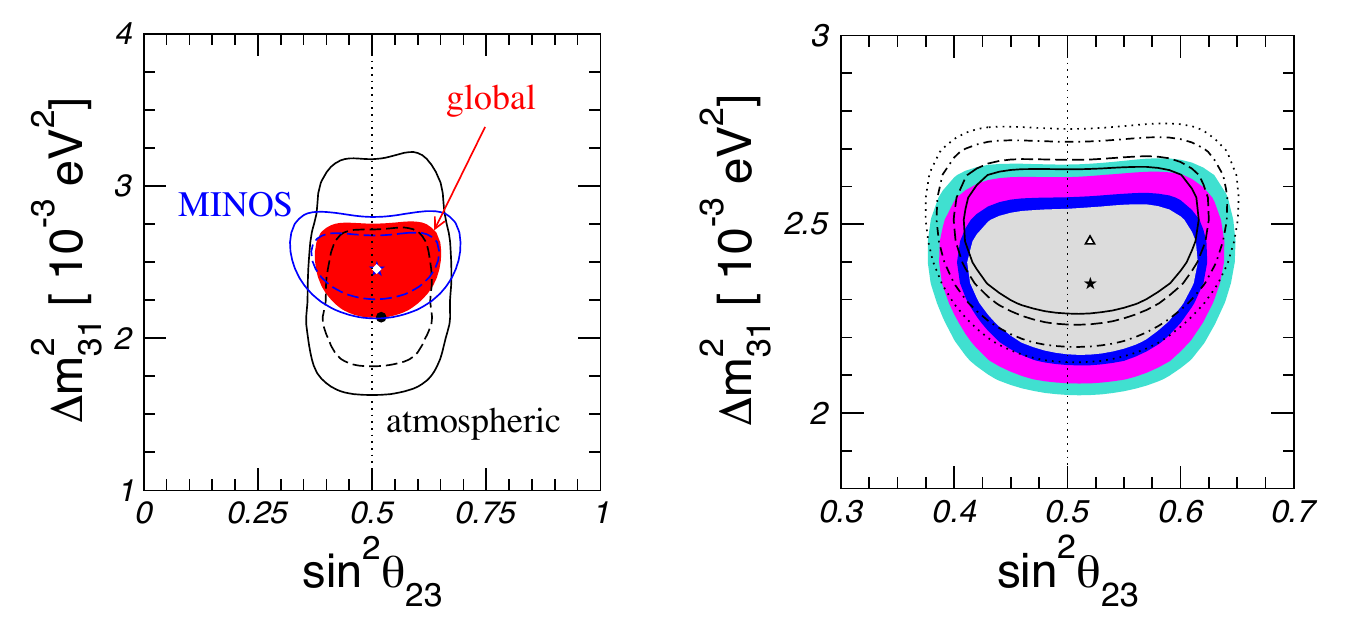}
\end{center}
\vspace{-0.7cm}
\caption{\label{fig:13sec}Determination of the atmospheric oscillation
  parameters. {\bf Left}: interplay of atmospheric (black) and MINOS
  disappearance (blue) data and the combination (red/shaded region)
  for normal hierarchy at 90\% C.L. (dashed) and $3\sigma$
  (solid). {\bf Right}: combined allowed regions for normal (black curves)
  and inverted hierarchy (colored regions) at 90\%, 95\%, 99\%, 99.73\%
  C.L.~\cite{Schwetz:2011qt}}
\end{figure}

The $|\Delta m^2_{31}|$ value is obtained from doing a combined fit of
MINOS and Super-Kamiokande atmospheric results. In addition the first
measurements of $\sin^2 2\theta_{23}$ using an off-axis neutrino beam
has been presented by the T2K collaboration~\cite{Abe:2012gx}. The
value of the mass splittings are well known but the ordering of the
masses, (the mass hierarchy), is still unknown. Discovering the mass
hierarchy is one of the goals of future experiments not only because
it obscures the CP violation measurement, but if neutrino masses were
inverted it would be the first time that fermions do not have
increasing masses with increasing generation number.

\section{Future plans}

With the discovery that $\theta_{13}$ is large, the gateway for
studying CP violation in the lepton sector and the mass hierarchy has
been opened. Future experiments plan to do just that. Three types of
proposals have emerged: high power conventional super beams,
beta-beams and neutrino factories. In addition given the large value
of $\theta_{13}$ Daya Bay could be able to study the mass hierarchy by
placing a 20~kton detector at 60~km~\cite{nuturn12,dayabay_nuturn12}.

\subsection{Super beams}
A super beam is a conventional beam, based on pion decays but reaching
powers of the order of the mega-watt. Three such beams have been
proposed. In Japan, the current T2K beam will be a super beam once it
reaches its design luminosity of 750kW. This beam in combination with
the proposed Hyper-Kamiokande detector~\cite{Abe:2011ts} would be a
very powerful tool to measure $\delta_{CP}$ assuming that the mass
hierarchy is known.  In the US the LBNE project plans to direct a
neutrino beam from FermiLab to the Homestake mine where a large LAr
detector would be built.  And in Europe the LAGUNA-LBNO project
presents three possible setups where the preferred option is a
super-beam from CERN-SPS to Pyh\"asalmi in Finland (baseline~=~2300~km).

Today's large value of $\theta_{13}$ implies that any future
experiment will quickly be systematic-limited. Therefore an effort
needs to be made to reduce systematic errors.  Experiments like
NA61~\cite{Abgrall:2011ae}, SciBoone~\cite{Mariani:2011xm},
MINER$\nu$A~\cite{MINERvA:2006aa} and nuSTORM~\cite{Kyberd:2012iz} are
extremely important to constrain hadron production cross-sections and
neutrino cross-sections.

\subsection{Beta-beams and neutrino factories}

The super beams can probably give us the value of $\delta_{CP}$ and
the mass hierarchy. But precision similar to the precision of the CKM
matrix will only be achieved with more powerful facilities. These
facilities will also be needed to test the unitarity of the PMNS
matrix and test whether neutrinos oscillate to become sterile
neutrinos. Beta-beams~\cite{Zucchelli:2001gp,Wildner:2011zza}, where a
radioactive ion is accelerated and then beta decays creating
$\bar{\nu}_e$ beam, could be an option.

Another option is the neutrino factory. Here, $\mu^-$ and $\mu^+$ trains
simultaneously circulate in opposite direction in the storage ring and
are separated in time. They are subsequently allowed to decay in
flight to create pure ($\nu_{\mu}$, $\bar{\nu}_e$)  and
($\bar{\nu}_{\mu}$, $\nu_e$) beams respectively. The golden channel of
the neutrino factory is $\nu_e \rightarrow \nu_{\mu}$, but since
$\nu_{\mu}$ are also present in the beam, measurements similar to
today's (electron neutrino appearance and muon neutrino disappearance)
are also possible. The presence of $\nu_{\mu}$, $\bar{\nu}_e$, and
$\bar{\nu}_{\mu}$, $\nu_e$ in different trains is perfect to study CP
violation since we can directly compare the oscillation of a neutrino
to the oscillation of the anti-neutrino.  Furthermore, because
$\nu_{\mu}$, $\bar{\nu}_e$, $\bar{\nu}_{\mu}$ and $\nu_e$ are present
in the beam, the near detector of a neutrino factory would be able to
measure all four types of neutrinos cross-sections and reduce
systematics considerably.  In order to see oscillations, the far
detector needs to be magnetized to differentiate $\nu_e \rightarrow
\nu_{\mu}$ from an interacting $\bar{\nu}_{\mu}$.  A magnetized iron
detector is envisaged as far detector. It was found that an optimal
neutrino factory given the current value of $\theta_{13}$ uses 10~GeV
muons and a baseline of 2200~km~\cite{ids-nf,huber_cern12}.

\section{Conclusions}

\begin{figure}[h]
\begin{center}
\includegraphics[width=25pc]{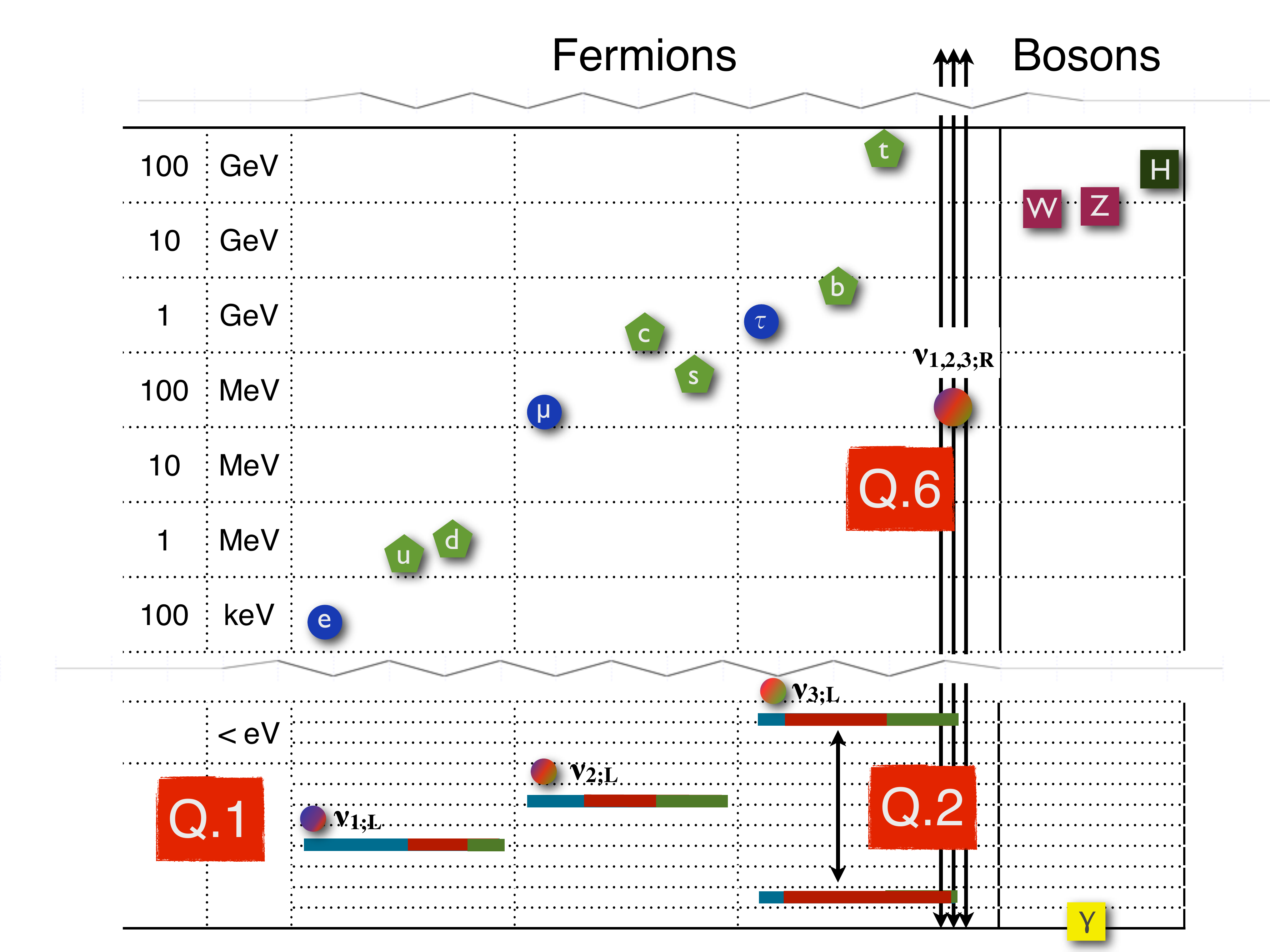}
\end{center}
\vspace{-0.5cm}
\caption{\label{fig:sm}The remaining questions of the Standard model
  are: {\bf Q.1} Determination of the absolute mass scale of neutrinos. {\bf Q.2}
  Determination of the mass hierarchy of the active neutrinos. {\bf Q.3} CP
  violation in neutrino oscillations. {\bf Q.4} Violation of unitarity of the
  neutrino mixing matrix. {\bf Q.5} Neutrinoless double beta decay. {\bf Q.6}
  Discovery of effects implying unambiguously the existence of sterile
  neutrino(s).}
\end{figure}

The years 2011 and 2012 have without any doubt seen tremendous
progress in neutrino physics. Only a year ago, nobody knew if
$\theta_{13}$ was even large enough to be measured, yet today the
precision on $\theta_{13}$ is already better than the precision on
$\theta_{23}$. With this measurement the prospect of measuring the
mass hierarchy and especially the CP phase $\delta$ is better than
ever and the years to come will be fascinating.  With the announcement
on July 4th 2012 that the Higgs was probably found, the remaining
questions in the Standard Model (Fig.~\ref{fig:sm}) are in neutrinos
physics.

We thank Alain Blondel and Mark Rayner for the careful reading of the
manuscript.

\section{References}
\bibliography{fdufour_biblio}

\end{document}